\documentclass{ws-procs975x65}
\usepackage{graphicx}
\usepackage{bm}
\usepackage{amsmath}
\usepackage[latin1]{inputenc}
\usepackage[english]{babel}
\usepackage[T1]{fontenc}
\usepackage{amssymb}
\usepackage{amsfonts}
\usepackage{epsfig}
\usepackage{colordvi}
\usepackage{psfrag}
\usepackage{color}
\newcommand{\beq}{\begin{equation}}
\newcommand{\eneq}{\end{equation}}

\begin{document}

\title{ $O(d,d)$ duality transformations in F(R) theories of gravity}

\author{ Gabriele Gionti, S.J.$^{1,2}$}
\address{$^1$ Specola Vaticana,
Vatican City, V-00120, Vatican City State and Vatican Observatory Research Group,
Steward Observatory, The University Of Arizona, 933 North Cherry Avenue
Tucson, Arizona 85721, USA.}

\address{ $^{2}$ Istituto Nazionale di Fisica Nucleare (INFN), Laboratori Nazionali di Frascati, Via E. Fermi 40, 00044 Frascati, Italy.} 

\begin{abstract}

The argument of Hodge duality symmetry is introduced starting from the electromagnetic field. Introducing bosonic string theory, $O(d,d)$ duality symmetry can be implemented when there exist $d$-symmetries, which allows one to write Hodge-dual fields. A tree-level effective gravitational action of bosonic string theory coupled with the dilaton field is considered. This theory inherits the Busher's duality of its parent string theory. The dilaton field can be recast into the Weyl's mode of the metric tensor in the Jordan frame. This maps the effective one-loop bosonic string  theory of gravity into a Lagrangian of a $f(R)$ function. Constraining this $f(R)$-Lagrangian on a FLRW metric and using Noether symmetries approach for extended theory of gravity, it is possible to show that the Lagrangian exibits a Gasperini-Veneziano duality symmetry. 
 
\end{abstract}

\keywords{ Modified theories of gravity, string theory,  scale factor duality.}


\bodymatter

\section{Introduction}
\label{uno}

The subject of Duality, in particular Hodge duality, has been fairly well known. If $F$ is the field strength in Electromagnetism

\beq
F\equiv F_{\mu\nu}dx^{\mu}\wedge dx^{\nu}\;\;\;\;,
\label{compoeltro}
\eneq

then the equations of motion,  in the Vacuum case, can be written in the following form 

\beq
dF=0\;\;\;\;\;\;\;\;\;d\star F=0\;\;,
\label{vuoto}
\eneq

\noindent  where $\star$ is the Hodge operator. Given a 2-form in four dimensions with a Lorentian signature $\star\star=-1$, it is easy to see that for the equations of motions $(\ref{vuoto})$, if $F$ is a solution also $\star F$, the Hodge dual of $F$, is a solution. This means that, naming $E$ and $B$ the electric and magnetic field respectively, if $(E,B)$ is a solution of the equation of motion then $( -B,E)$ is also a solution. This invariance of the equation of motion is called duality symmetry of electromagnetism. In fact, the electromagnetic field in the vacuum can be described both by the Lagrangian density 

\beq
L =-\frac{1}{4} F \wedge {\star F} \;\;\;\;,
\label{laga}
\eneq

\noindent and the Lagrangian density of it related Hodge-dual 2-form $\star F$

\beq
{\tilde L} =\frac{1}{4} {\star F }\wedge F \;\;\;\;,
\label{dilaga}
\eneq

\noindent since the two Lagrangians differs, as may be easily seen in componets of $(E,B)$, by a $-$ sign, but have the same equations of motions (\ref{laga}) \cite{Zwiebach}. 

\section{Duality in bosonic string theory}
\label{due}

Analogous considerations can be done in the case it is studied Hodge-duality in bosonic string theory. 



In fact, consider a two-dimensional sigma model defined on a worldsheet, that is a two-dimensional Riemannian manifold $(M,h)$ embedded in a $(n+1)$ Lorentzian Manifold $\{{\cal M},G\}$.  The action is (see Ref.\refcite{busher1} for all definitions) 

\begin{equation}
S=\frac{1} {4 \pi {\alpha}'} \int d^{2}\xi  \Big (\sqrt{h}h^{\mu\nu}G_{ab}(\partial_{\mu}X^{a})(\partial_{\nu}X^{b})+\epsilon^{\mu\nu}B_{ab}(\partial_{\mu}X^{a})(\partial_{\nu}X^{b})
+{\alpha}' {\sqrt{h}}R^{(2)}\phi(X^{a})\Big),
\label{totstring}
\end{equation}


The dual Lagrangian ${\tilde {\cal L}}$ can be defined if there exists a vector field $Y$ which is an isometry for ${\cal L}$\cite{Rocek:1991ps,Alvarez:1994dn}, that is  $ L_{X} { \cal L}$, where $L_{X}$  is the Lie derivative along a vector filed $X$ of ${\cal L}$. In  the case ${\displaystyle Y=\frac{\partial}{\partial X^{0}}}$, the dual action is 
\begin{equation}
{\tilde S}=\frac{1} {4 \pi {\alpha}'} \int d^{2}\xi  \Big (\sqrt{h}h^{\mu\nu}{\tilde G}^{ab}(\partial_{\mu}{\tilde X}_{a})(\partial_{\nu}{\tilde X}_{b})+\epsilon^{\mu\nu}{\tilde B}^{ab}(\partial_{\mu}{\tilde X}_{a})(\partial_{\nu}{\tilde X}_{b})
+{\alpha}' {\sqrt{h}}R^{(2)}\phi(X^{a})\Big),
\label{duatotstring}
\end{equation}
where 
\begin{equation}
{\tilde G}_{00}=1/G_{00}\,, \,\,\,\, {\tilde G}_{0i}=B_{01}/G_{00}\,, \,\,\,\, {\tilde G}_{ij}=G_{ij} -  (G_{0i}G_{0j}-B_{0i}B_{0j})/G_{00}\,, \,\,\,\, i,j=1,...n\,.
\label{medual}
\end{equation}
and 
\begin{equation}
{\tilde B}_{0i}=-{\tilde B}_{i0}=G_{0i}/G_{00}\,, \,\,\,\, {\tilde B}_{ij}=B_{ij}+(G_{0i}B_{0j}-B_{0i}G_{0j})/G_{00}\,, \,\,\,\, i,j=1,...,n\,.
\label{macca}
\end{equation}
Eqs. (\ref{medual}) and (\ref{macca}) are the so called  {\it Buscher duality relations} for this particular isometry~\cite{busher1,busher2}. The dual action is dynamically equivalent to the original one at classical level.

Very often it is assumed that the background fileds $G$ and $B$ are constant in the action (\ref{totstring}). 









 The Hamiltonian density $H$ is expressed \cite{Maharana,Francocorfu,R} as
\beq
H=\frac{1}{2}Z^T M(G,B)Z \;\;,
\eneq
where
\beq
Z=\begin{pmatrix} P \cr X' \cr\end{pmatrix}\;\;,
\eneq
and it has been suppressed the indices; P is the conjugate momentum, while $X'$ the derivative of $X$ respect to the spatial coordinate on the worldsheet, while M is the following matrix:
\beq
 M=\begin{pmatrix}G^{-1} & -G^{-1}B \cr
             BG^{-1} & G-BG^{-1}B \cr\end{pmatrix}\;\;,
\eneq
is a symmetric $(n+1)\times (n+1)$ matrix.

Note that under interchange
$ P\leftrightarrow X'$, the Hamiltonian density remains invariant if we
simultaneously transform $ M\leftrightarrow  M^{-1} $.
The Hamiltonian density is also invariant under the following global
 $ O(n+1, n+1) $ transformations:
 The $Z$-vector and  $ M$-matrix transform as
\beq
Z\rightarrow \Omega Z,~~ M\rightarrow\Omega M\Omega ^T,~
\eta=\Omega\eta\Omega^T,~~\Omega\in O(n+1, n+1)\;\;,
\eneq
 where  $ \eta$  is the  $O(n+1, n+1)$ metric.
\beq
\eta=\begin{pmatrix} 0 &  1 \cr 1 & 0 \cr\end{pmatrix}\;\;,
\eneq
where  $1$ is  $ n+1\times n+1$ unit matrix. $Z$ is
$2n+2$-dimensional $O(n+1,n+1)$ vector. This shows that the duality symmetry group ($G$ and $B$ constant)  is $O(n+1,n+1)$.

 \section{Gasperini-Veneziano duality}
 \label{tre}
 As it is well known, a two dimensional sigma model is Weyl's invariant at classical level. Requiring the Weyl's invariance at quantum level as well, that means that the path-integral associate to the action (\ref{totstring}) has to be invariant, this implies that the equations deriving form the variations of the following action $S^{n+1}$  ~\cite{busher1},

\begin{equation}
S^{n+1}=K\int d^{n+1} x \sqrt{-G}e^{-2\phi}\left (R+4{\nabla}_{a}\phi{\nabla}^{a}\phi - {\frac{1}{12}}H^{2} + \Lambda\right)\,,
\label{effetto}
\end{equation} 
where $H_{abc}=\partial_{a}B_{bc}+$ {\it cyclic permutations} and $\nabla$ is the covariant derivative on the target space, need to be satisfied. The cosmological constant $\Lambda$ can be defined as  ${\displaystyle \Lambda \equiv {\frac{1}{{\alpha}'}}{\frac{n-25}{48{\pi}^{2}}}}$, which is equal to $0$ if and only if $n=25$ (the critical dimension for bosonic String Theory). 

Once the conditions derived from the action (\ref{effetto}) are satisfied, then also the dualized model (\ref{duatotstring}) is conformal invariant at one loop, provided that the dilaton field shifts in the following way~\cite{busher1}:
\begin{equation}
{\tilde \phi}=\phi -\frac{1}{2}\ln {(g_{00})}.
\label{conf}
\end{equation}

Meissner and Veneziano have shown \cite{veneziano2,Meissner} that the action (\ref{effetto}), in case $G$ and $B$ are at most time dependent, is $O(n,n)$ invariant. 
Gasperini and Veneziano~\cite{gasperini1} have shown that if the following spatially flat, homogeneous and isotropic metric in $n+1$ dimensions
\begin{equation}
ds^{2}=dt^{2}-a^{2}(t)dx^{2}_{i}\,,
\label{metro}
\end{equation}

where $a(t)$ is the scale factor, is a solution of the equations of motions derived from (\ref{effetto}), then also the metric with $a(t) \rightarrow a^{-1}(\pm t)$ is a solution. The effective gravitational action (\ref{effetto}) exhibits the following duality symmetric transformations among the solutions of its Euler-Lagrange equations (which are nothing else but an application of the Busher's duality relations in the case of a symmetry made out by $n$-Abelian spatial  vector fields ${\displaystyle \frac {\partial}{\partial x^{i}}}$)
\begin{equation}
a(t) \rightarrow a^{-1}(\pm t), \,\,\,\, \phi \rightarrow {\tilde \phi}=\phi-n \ln(a)\,.
\label{cosmdual}
\end{equation} 
This is the  {\it long-short}  duality correspondence of the scale factor $a$, which is a particular case of a $O(n,n)$ transformation, for the string-dilaton cosmology. These duality relations (\ref{cosmdual}) among the solutions of the Euler-Lagrange equations of (\ref{effetto}) allow  to construct the so called {\it Pre-Big Bang} cosmological models~\cite{gasperini1}.

\section{Duality in cosmology of modified gravity}
\label{cinque}


Consider now the \textit{tree level dilaton-graviton string effective action} 
\beq
\mathcal{S}=\int d^Dx\sqrt{-G}e^{-2\phi}\left[R+4\nabla_\mu\phi\nabla^\mu\phi+\Lambda\right],   \label{2.0}
\eneq
that can be achieved from the above results, in the low-energy limit, retaining only the scalar mode (the dilaton) and the tensor mode (the graviton)~\cite{veneziano1, gasperini1}. Here $D=n+1$ is the number of the dimensions (spatial + time), and $G$ indicates the determinant of the $D$-dimensional spacetime metric.
In this sense, the  effective string action  reduces to a scalar-tensor theory (Brans-Dicke-like). 
 
From now on, it is considered the case in which the space-time has $D=4$ dimensions, so that the String is not in critical dimensions and $\Lambda \neq0$. The main idea is to explore the possibility of writing the action (\ref{2.0}) as a $f(R)$-modified theory of gravity. 
The string-dilaton effective Lagrangian can be connected to the $f(R)$ gravity Lagrangian via the Weyl's transformation (see Ref.\refcite{gabjcap}). 
 In fact the two actions can be mapped into each other as 
\begin{equation}
\sqrt{-g}e^{-2\phi}\left(R+4\nabla_\mu\phi\nabla^\mu\phi+\Lambda \right)=\sqrt{-\tilde{g}}f(\tilde{R})\,,
\end{equation}
Using some relations on the equation of motions as   in \cite{gabjcap}, it is possible to arrive to 
\beq
f(\tilde{R})=2\Lambda e^{2\phi}\,.     \label{3.23}
\eneq

Consider, now, a homogeneous and isotropic universe described by FLRW metric in spherical coordinates, 
\beq
ds^2=dt^2-a^2(t)\left[\frac{dr^2}{1-kr^2}+r^2\left(d\theta^2 +\mbox{sin}^2 \theta d\phi^2\right)\right],
\eneq

$a(t)$ being the scale factor  and $k$ is the spatial curvature. 
In order to simplify the problem, it is chosen $k=0$. Then Ricci's trace $R$ is 
\beq
R=-6\left[\frac{\ddot{a}}{a}+\left(\frac{\dot{a}}{a}\right)^2\right].            \label{4.14} 
\eneq
The Lagrangian of this cosmological model is $\mathcal{L}\equiv\mathcal{L}(a,\dot{a},R,\dot{R})$, where the configuration space is $\mathcal{Q}=\left\{a,R\right\}$ and $\mathcal{TQ}=\left\{a,\dot{a},R,\dot{R}\right\}$ is the related tangent bundle on which $\mathcal{L}$ is defined.
The corresponding point-like action is \cite{mauro,cimall,PhysRept}
\beq
S=\int \mathcal{L}(a,\dot{a},R,\dot{R})dt,                                  
\eneq
where $a$ and $R$ are independent variables. A Lagrange multiplier is introduced such that Eq.(\ref{4.14}) holds along the equations of motion.
The above action can be recast as
\begin{equation}\label{10}
S=4\pi\int dt \left\{ a^3f(R)-\lambda\left [ R+6\left (
\frac{\ddot{a}}{a}+\frac{\dot{a}^2}{a^2}\right)\right]\right\}\,{,}
\end{equation}
where the Lagrange multiplier $\lambda$ can be obtained by varying with respect to the Ricci scalar $R$, giving
\begin{equation}\label{11}
\lambda=a^3f'(R)\,.
\end{equation}

Using Noether's symmetries approach, it is possible to show that there exists a class of $f(R)$ functions for which \cite{gabjcap}

\beq
f''(R)\dot{R}=f'(R)\left(\frac{\dot{a}}{a}\right)\,                                \label{5.2}\;\;\;\;.
\eneq
 A simplified Lagrangian is obtained 
\beq
\mathcal{L}=a^3\left[f(R)-f'(R)R\right]+12a\dot{a}^2f'(R).           \label{5.3}
\eneq  



Being  $a^{3}=e^{3log(a)}$, the previous Lagrangian can be rewritten as

 \begin{equation} 
\mathcal{L}=\left[(2\Lambda e^{2(\phi +\frac{3}{2}log(a))}-4\Lambda R e^{2(\phi +\frac{3}{2}log(a))}{\phi}'(R))+48\left(\frac{\dot{a}}{a}\right)^{2} \Lambda e^{2(\phi +\frac{3}{2}log(a))}\phi'(R) \right]\,. 
\label{resco2}
\end{equation}
Applying the dilation shift as similarly done in Eqs. (\ref{conf}) and (\ref{cosmdual}), the dilaton can be redifined as  
\begin{equation}
\phi \mapsto {\tilde\phi}=\phi + \frac{3}{2}\ln (a)\,,
\label{sco}
\end{equation}

\noindent therefore eq. (\ref{resco2}) becomes:

\begin{equation} 
\mathcal{L}=\left[2\Lambda e^{2{\tilde\phi}}-4\Lambda R e^{2{\tilde \phi}}{\tilde \phi}'(R)\right]+48\left(\frac{\dot{a}}{a}\right)^{2} \Lambda e^{2{\tilde \phi}}{\tilde\phi}'(R)\,, 
\label{resco3}
\end{equation}
and then reapplying equation (\ref{3.23})
\beq
\mathcal{L}=\left[f(R)-f'(R)R\right]+12\left(\frac{\dot{a}}{a}\right)^2f'(R)\,,                     \label{5.4}
\eneq
the scale factor duality $a\rightarrow 1/a$ holds.  

\section{Conclusion}
\label{sei}

In this short essay, the issue of duality symmetry, for some particular modified theories of gravity, has been analyzed. 
The connections, if any, between $O(d,d)$ duality symmetry and Noether's symmetry approach could be a further possible developments of this research. More precisely, the fact  that the action of tree level effective string theory of gravity is, in the case either $B$ and $G$ are constant or function of the time only, $O(d,d)$ invariant, the question of how this symmetry is linked to Noether's symmetry approaches follows naturally. In fact, Noether's symmetry approach to alternative theory of gravity could be a useful tool to select $f(R)$ theories of gravity, which exhibit explicitly duality symmetries.

\end{document}